\documentclass[twocolumn,showpacs,preprintnumbers,amsmath,amssymb]{revtex4-1}
\usepackage{graphicx}
\usepackage{dcolumn} 
\usepackage{bm}      
\usepackage{xcolor}      
\usepackage[mathscr,mathcal]{euscript}

\begin{document}

\title{Topological non-symmorphic crystalline insulators}

\author{Chao-Xing Liu$^1$ and Rui-Xing Zhang$^1$}
\affiliation{$^1$Department of Physics, The Pennsylvania State University, University Park, Pennsylvania 16802-6300}


\date{\today}

\begin{abstract}
In this work, we identify a new class of $Z_2$ topological insulator protected by non-symmorphic crystalline symmetry, dubbed a ``topological non-symmorphic crystalline insulator''. We construct a concrete tight-binding model with the non-symmorphic space group pmg and confirm the topological nature of this model by calculating topological surface states and defining a $Z_2$ topological invariant. Based on the projective representation theory, we extend our discussion to other non-symmorphic space groups that allows to host topological non-symmorphic crystalline insulators.
\end{abstract}

 \maketitle

{\it Introduction -}
Search for new states of matters, especially those with non-trivial topological properties, is one of the main focuses in condensed matter physics. Recent experimental discovery of time reversal (TR) invariant topological insulators (TIs)\cite{qi2010phystoday,hasan2010,moore2010,qi2011} has inspired lots of research interests and led to a rapid development in this field. TR invariant TIs possess insulating bulk states and metallic edge/surface states that are protected by TR symmetry due to the double degeneracy by Kramers' theorem\cite{kane2005b}. In principle, degeneracies can also come from other types of symmetries, such as crystalline symmetry, particle-hole symmetry, {\it etc}. Therefore, it is natural to ask if one can find new topological phases protected by other symmetries. Several recent theoretical works\cite{fu2011a,hsieh2012,slager2012,fang2012a,jadaun2012,fang2012b,kargarian2013,ueno2013,tanaka2013,liu2013d,fang2012c} are devoted to topological phases protected by crystalline symmetry, dubbed ``topological crystalline insulators (TCIs)''. Recent experimental observations of surface states in materials consisted of SnTe confirm that this system is a TCI protected by mirror symmetry\cite{dziawa2012,tanaka2012,xu2012a}.

Non-symmorphic crystals possess symmetry operations combining a point symmetry operation and a non-primitive translation operation, which cannot be removed by choosing the origin of crystals\cite{dresselhaus2008,bir1974}. It is known that non-symmorphic symmetries can ``stick bands together'' and yield extra degeneracies. Most studies of TCIs have been focusing on point groups or symmorphic part of space groups and it is still unclear if non-symmorphic symmetries can yield new TCIs. In this letter, we give an affirmative answer to this question by constructing a concrete example and identify a new class of $Z_2$ topological phase as a direct physical consequence of non-commutativity of symmetry operators in non-symmorphic groups. Consequently, we name it as a ``topological non-symmorphic crystalline insulator'' (TNCI). Our general discussion on non-symmorphic space groups will provide a guidance to search for realistic materials with the TNCI phase.

\begin{figure}
   \begin{center}
      \includegraphics[width=3.5in,angle=0]{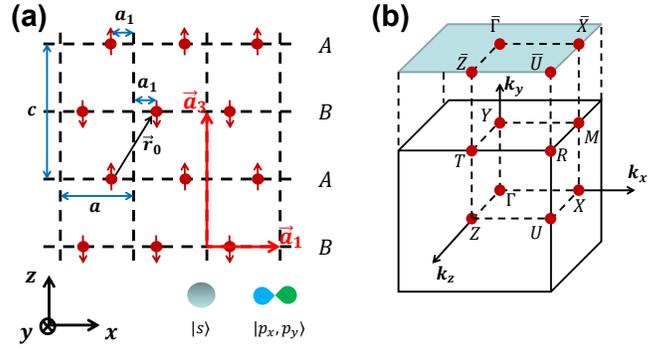}
    \end{center}
    \caption{ (a) Lattice sites in the xz plane for our tight-binding model. The lattice vectors $\vec{a}_1$ and $\vec{a}_3$ are shown in the figure and $\vec{a}_2$ is perpendicular to this plane (y direction). (b) The bulk Brillouin zone and the surface Brillouin zone for the xz plane.     }
    \label{fig1}
\end{figure}

{\it Tight-binding model -}
As shown in Fig. \ref{fig1}(a), our tight-binding model has a layered antiferromagnetic structure stacked along z direction, where each layer is a square lattice with magnetic moments on each site perpendicular to the xy plane. Magnetic moments in each layer are polarized along the same direction, so they are ordered ferromagnetically. However, the magnetization directions between two adjacent layers are opposite, so the whole system has an antiferromagnetic structure with two atoms in one unit cell, denoted as A and B. The lattice vectors are denoted as $\vec{a}_1 = (a, 0, 0)$, $\vec{a}_2 = (0, a, 0)$ and $\vec{a}_3=(0,0,c)$. In a unit-cell, A and B atoms are shifted in opposite directions along x axis. The position of A atom is $r_A=(-a_1,0,0)$ while that of B atom is $r_B=(a_1,0,\frac{c}{2})$, as shown in Fig. \ref{fig1}(a). Each lattice site contains three orbitals $|s\rangle$, $|p_x\rangle$ and $|p_y\rangle$. The $|p_x\rangle$ and $|p_y\rangle$ orbitals carry the angular momentum 1 and couple to magnetic moments through Zeeman type of coupling, denoted as $M_1$. The explicit form of our Hamiltonian are shown in supplementary materials\cite{liu2013note}, which is written under the basis
$|\alpha\eta,\vec{k}\rangle=\frac{1}{\sqrt{N}}\sum_n e^{i\vec{k}\cdot \vec{r}_{n\eta}}\phi_{\alpha}(\vec{r}-\vec{r}_{n\eta})$
where $N$ is the normalization factor, $\vec{r}_{n\eta}=\vec{R}_n+\vec{r}_{\eta}$ with the lattice vector $\vec{R}_n$ and the position $\vec{r}_{\eta}$ of the atom $\eta=A,B$, and $\phi_\alpha$ denotes the basis wavefunction ($\alpha=s,p_x,p_y$).

Our model is quite similar to that discussed previously by one of the authors\cite{liu2013c}, where the anti-unitary operation combining TR symmetry with translation symmetry plays an essential role. However, in the present model, the shifting of A and B atoms in opposite directions breaks this anti-unitary symmetry. Instead, it turns out that two unitary operators are essential. One is the mirror symmetry along z direction, denoted as $\hat{m}_z=\{\hat{m}_z|\vec{e}\}: (x,y,z)\rightarrow (x,y,-z)$ where $\vec{e}=(0,0,0)$, and the other is the glide symmetry $\hat{g}_x=\{\hat{m}_x|\vec{\tau}\}: (x,y,z)\rightarrow (-x,y,z+\frac{c}{2})$ with $\vec{\tau}=\frac{\vec{a}_3}{2}=(0,0,\frac{c}{2})$. These two symmetry operations, together with translation symmetry, give the 2D non-symmorphic space group pmg. Direct calculation gives
\begin{eqnarray}
\hat{m}_z \hat{g}_x=\{C_{2y}|\vec{\tau}\}\neq \hat{g}_x\hat{m}_z=\{C_{2y}|-\vec{\tau}\},
	\label{eq:mzg}
\end{eqnarray}
where $C_{2y}$ is a two-fold rotation around y axis. $\hat{m}_z \hat{g}_x$ is different from $\hat{g}_x\hat{m}_z$ by a shift of the primitive lattice vector $2\vec{\tau}=\vec{a}_3$. The non-commutativity between $\hat{g}_x$ and $\hat{m}_z$ is essential, as discussed below.

Let us first analyze the symmetry properties of our tight-binding Hamiltonian. On the basis $|\alpha\eta,\vec{k}\rangle$, the symmetry operators $\hat{m}_z$ and $\hat{g}_x$ behave as $\hat{m}_z|\alpha\eta,\vec{k}\rangle=|\alpha\eta,\hat{m}_z\vec{k}\rangle$ and $\hat{g}_x|\alpha\eta,\vec{k}\rangle=\sum_{\beta}e^{-i\pi k_z}m_{x,\alpha\beta}|\beta\bar{\eta},\hat{m}_x\vec{k}\rangle$
where $\bar{\eta}$ means the interchange of the A and B indices and the $3\times3$ matrix $m_{x}=\mbox{Diag}[1,-1,1]$ in the basis $|s\rangle,|p_x\rangle,|p_y\rangle$.
For a symmetry operation $\hat{U}$, the Hamiltonian should satisfy $H(\vec{k})=U^*(\vec{k})H(\hat{U}\vec{k})U^T(\vec{k})$. The details about how these symmetries constrain the form of Hamiltonian are shown in the supplementary material\cite{liu2013note} and we focus on the $k_z=\frac{\pi}{c}$ plane here. Since $|\alpha\eta,\vec{k}+\vec{G}\rangle=e^{i\vec{G}\cdot\vec{r}_{\eta}}|\alpha\eta,\vec{k}\rangle$, the off-diagonal part Hamiltonian $H_{AB}(\vec{k})$ is not periodic, but satisfies the relation $H_{AB}(\vec{k}+\vec{G})=e^{i\vec{G}\cdot\vec{r}_{0}}H_{AB}(\vec{k})$ where $\vec{r}_0=\vec{r}_B-\vec{r}_A$. At $k_z=\pi/c$, one has $H_{AB}(k_x,k_y,\frac{\pi}{c})=-H_{AB}(k_x,k_y,-\frac{\pi}{c})$.
But due to mirror symmetry $\hat{m}_z$, $H_{AB}(k_x,k_y,\frac{\pi}{c})=H_{AB}(k_x,k_y,-\frac{\pi}{c})$, so $H_{AB}(k_x,k_y,\frac{\pi}{c})=0$, which means that there is no coupling between the A and B layers and the Hamiltonian is block-diagonal in the $k_z=\pi/c$ plane. We will always denote the momentum in the $k_z=\pi/c$ plane by $\vec{\kappa}=(k_x,k_y,\frac{\pi}{c})$ below. If $|\phi_{A,\vec{\kappa}}\rangle$ is an eigenstate of $H_A(\vec{\kappa})$ with eigenenergy $E_{A,\vec{\kappa}}$,
$|\phi_{B,\hat{m}_x\vec{\kappa}}\rangle=\hat{g}_x|\phi_{A,\vec{\kappa}}\rangle$ is an eigenstate of $H_B(\hat{m}_x\vec{\kappa})$ with the same eigenenergy ($H_B(\hat{m}_x\vec{\kappa})|\phi_{B,\hat{m}_x\vec{\kappa}}\rangle=E_{A,\vec{\kappa}}|\phi_{B,\hat{m}_x\vec{\kappa}}\rangle$). Therefore, we find $|\phi_{A,\vec{\kappa}}\rangle$ is degenerate with $|\phi_{B,\hat{m}_x\vec{\kappa}}\rangle$ at the $k_z=\frac{\pi}{c}$ plane. When $\vec{\kappa}$ satisfies $\hat{m}_x\vec{\kappa}=\vec{\kappa}$, which is true for two lines given by $(0,k_y,\frac{\pi}{c})$ and $(\frac{\pi}{a},k_y,\frac{\pi}{c})$ in momentum space, all the electronic states are doubly degenerate.

\begin{figure}
   \begin{center}
      \includegraphics[width=3.5in,angle=0]{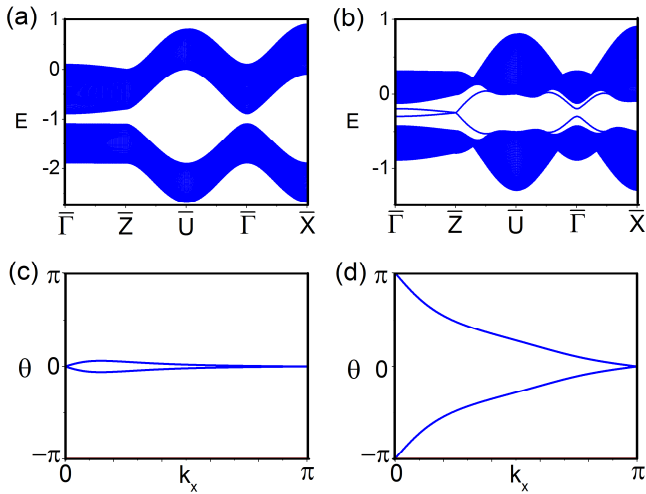}
    \end{center}
    \caption{ Energy dispersion of a slab configuration for our tight-binding model with the parameter (a) $M_1=3.1$ and (b) $M_1=4.5$. The corresponding evolution of Wannier function centers in a 3D bulk system is shown as a function of $k_x$ for (c) $M_1=3.1$ and (d) $M_1=4.5$. The surface states in (b) and the winding number of Wannier function centers in (d) indicate that the system is in the TNCI phase for $M_1=4.5$. }
    \label{fig2}
\end{figure}

To confirm the existence of topological phases in our model, we perform an electronic structure calculation for a slab configuration with finite lattice sites along the y direction. Since the surface of the slab is normal to the y direction, the symmetries $\hat{m}_z$ and $\hat{g}_x$ are preserved. We find that when the coupling $M_1$ between magnetization and p-orbitals is small, there are no surface states (Fig. \ref{fig2}(a)). But when $M_1$ exceeds a critical value, Dirac type of surface states appear around $\bar{Z}$, as shown in Fig. \ref{fig2}(b). The degeneracy of Dirac point at $\bar{Z}$ is due to two unitary symmetry operators $\hat{m}_z$ and $\hat{g}_x$, as shown above. It is impossible to remove surface states in Fig. \ref{fig2}(b) without closing bulk band gap as long as pmg group symmetry is preserved, so we expect that a topological phase protected by the pmg group exists in this system. In the following, we will define a bulk topological invariant for this topological phase.

{\it $Z_2$ topological invariant -}
To confirm the topological nature of two dimensional (2D) surface states, it is necessary to construct a topological invariant for the three dimensional (3D) bulk system. The $Z_2$ topological invariant in TR invariant TIs is defined by the Pfaffian of the asymmetric matrix of TR operator in the occupied band subspace\cite{fu2007a,kane2005b}. As discussed above, only unitary symmetry operators are involved in our model, so there is no anti-unitary operator to replace the TR operator to define topological invariants. Nevertheless, we can still separate all the occupied states into two sets and introduce the concept of ``partial polarization'' for each set\cite{fu2007a}. The $Z_2$ topological invariant can be defined by tracking the evolution of this partial polarization.

We start from identifying two sets of degenerate eigenstates for a generic system with the pmg group symmetry. Since the Hamiltonian has $\hat{m}_z$ symmetry, one can find the common eigenstates of $H(\vec{k})$ and $\hat{m}_z$ at the $k_z=0$ and $k_z=\pi/c$ plane. We consider the $k_z=\pi/c$ plane and take one common eigenstate $|\phi^{I}_{\vec{\kappa}}\rangle$ given by $H(\vec{\kappa})|\phi^I_{\vec{\kappa}}\rangle=E_{I,\vec{\kappa}}|\phi^I_{\vec{\kappa}}\rangle$ and $\hat{m}_z|\phi^I_{\vec{\kappa}}\rangle=m_z|\phi^I_{\vec{\kappa}}\rangle$. One can define a state
$|\phi^{II}_{\hat{m}_x\vec{\kappa}}\rangle=e^{i\chi_{\kappa}}\hat{g}_x|\phi^{I}_{\vec{\kappa}}\rangle$,
which is an eigenstate of $H(\hat{m}_x\vec{\kappa})$ with the same eigenenergy $E_{I,\vec{\kappa}}$, but acquires a phase shift $\chi_{\vec{\kappa}}$. Moreover, according to (\ref{eq:mzg}), direct calculation shows $\hat{m}_z\hat{g}_x|\phi^{\alpha}_{\vec{\kappa}}\rangle=iC_{2y}|\phi^{\alpha}_{\vec{\kappa}}\rangle$ and $\hat{g}_x\hat{m}_z|\phi^{\alpha}_{\vec{\kappa}}\rangle=-iC_{2y}|\phi^{\alpha}_{\vec{\kappa}}\rangle$ ($\alpha=I,II$), so $\{\hat{m}_z,\hat{g}_x\}=0$ at the $k_z=\pi/c$ plane, which indicates that the mirror parity of $|\phi^{II}_{\hat{m}_x\vec{\kappa}}\rangle$ is $-m_z$, different from that of $|\phi^I_{\vec{\kappa}}\rangle$. Therefore, one finds two distinct sets of eigenstates, dubbed ``doublet pairs'' below.

With doublet pairs, we can define the partial polarization as $P_{\alpha}(k_x)=\frac{1}{2\pi}\oint dk_y\langle\phi^{\alpha}_{\vec{\kappa}}|i\partial_{k_y}|\phi^{\alpha}_{\vec{\kappa}}\rangle$ ($\alpha=I,II$). The partial polarizations of doublet pairs can be related to each other by
\begin{eqnarray}
	P_{II}(-k_x)=P_{I}(k_x)-\frac{1}{2\pi}(\chi_{\pi/a}-\chi_{-\pi/a}).
	\label{eq:Polarization}
\end{eqnarray}
Due to the single-valueness of $|\phi^{\alpha}_{\vec{\kappa}}\rangle$, the phase $\chi_{\vec{\kappa}}$ can only differ by $2\pi$ times an integer when $k_x$ is changed by $2\pi/a$. Thus, Eq. (\ref{eq:Polarization}) leads to two conclusions: (1) $P_{II}$ at $k_x\in[-\pi/a,0]$ is determined by $P_I$ at $k_x\in[0,\pi/a]$; (2) at $k_x=0$ and $\pi/a$, $P_I$ is equivalent to $P_{II}$ up to an integer.

The constraint of the parital polarization from Eq. (\ref{eq:Polarization}) indicates the possibility of defining a $Z_2$ topological invariant. Based on the method introduced by Yu {\it et. al.}\cite{yu2011}, one can obtain the Wannier function centers $\theta$ of the occupied bands by calculating the eigenvalues of non-Abelian Berry connection along the ``Wilson loop''. The polarization is related to Wannier function centers by $P=\frac{\theta}{2\pi}$. The Wannier function centers of doublet pairs as a function of $k_x$ are shown in Fig. \ref{fig2} (c) and (d). One can clearly see the different evolutions of Wannier function centers between topologically trivial and non-trivial phases. Wannier function centers are periodic and only well-defined by any integer times $2\pi$. Thus, one can regard the regime $[-\pi,\pi]$ as a ring and consider the evolution of Wannier function centers on this ring. Similar to the case of TR invariant TIs\cite{yu2011}, the total winding number of the Wannier function centers of all doublet pairs on this ring defines a $Z_2$ topological invariant. pmg symmetry group guarantees that the Wannier function centers of doublet pairs must be degenerate at $k_x=0$ and $k_x=\pi/a$. If the Wannier function centers of all doublet pairs enclose the ring an odd number of times, it is topologically non-trivial. Otherwise, it is topologically trivial. Alternatively, one can define the ``doublet polarization'' $P_d=P_I-P_{II}$, in analog to the ``time reversal polarization'' introduced by Fu and Kane\cite{fu2007a}, and the $Z_2$ topological invariant is defined by the difference
$\Delta=P_d(\pi)-P_d(0)\mbox{  mod  }2$.

\begin{figure}
   \begin{center}
      \includegraphics[width=3.5in,angle=0]{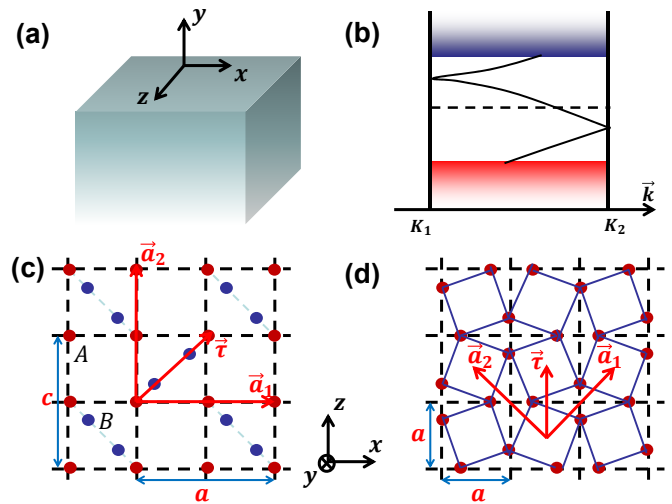}
    \end{center}
    \caption{ (a) Schematic plot of a semi-infinite crystal with one surface. (b) Energy dispersion of non-trivial surface states. Here $\vec{K}_1$ and $\vec{K}_2$ are two HSM. (c) A typical lattice structure with pgg group symmetry. Two lattice vectors are $\vec{a}_1=(a,0,0)$ and $\vec{a}_2=(0,0,c)$, and the vector $\vec{\tau}=(\frac{a}{2},0,\frac{c}{2})$. A and B atoms are denoted by red and blue colors. Two generators of the pgg group are $\{\hat{m}_x|\vec{\tau}\}$ and $\{\hat{m}_z|\vec{\tau}\}$. (d) A typical lattice structure with p4g group symmetry. The lattice vectors $\vec{a}_1=(a,0,a)$ and $\vec{a}_2=(-a,0,a)$. The vector $\vec{\tau}=(a,0,0)$. Two generators of p4g group are $\{C_4|\vec{e}\}$ and $\{\hat{m}_x|\vec{\tau}\}$.     }

    \label{fig3}
\end{figure}

{\it Other non-symmorphic groups - }
We will generalize our discussion to other non-symmorphic symmetry groups. From the above model, we can see that the degeneracies guaranteed by non-symmorphic symmetry plays an essential role in protecting surface states. However, the symmetry groups of different surfaces are different and not all the 2D surfaces can possess non-symmorphic symmetry even for a 3D non-symmorphic crystal. Therefore, our strategy is to directly consider a semi-infinite crystal with one specific surface, as shown in Fig. \ref{fig3}(a), of which the symmetry group can be described by a 2D space group (also known as a wallpaper group). We consider an insulating material in this semi-infinite configuration, and assume that the states are doubly degenerate at two high symmetry momenta (HSM) $K_1$ and $K_2$ in the surface Brillouin zone (BZ). As shown in Fig. \ref{fig3}(b), if surface states switch their degenerate partners between $K_1$ and $K_2$, such surface states cannot be adiabatically connected to any trivial state in a 2D system with the same symmetry group. Due to the boundary-bulk correspondence, we expect that topological phases can exist in the corresponding 3D bulk systems. In 2D, there are only 16 wallpaper groups, so a systematic study of TCIs is possible. We focus on non-symmorphic symmetry groups here.

The above analysis has shown the importance of symmetry induced degeneracy. The degeneracy of electronic states in a system can be determined by high dimensional irreducible representations (IRs) of its symmetry group\cite{dresselhaus2008,bir1974}. In a space group, symmetry operations at a fixed momentum usually form a subgroup of the whole group, known as a wavevector group or a little group. Consequently, the degeneracy of electronic states at a certain momentum is determined by the IRs of the wavevector group. For the 2D non-symmorphic groups considered here, it turns out that only four HSM $\bar{\Gamma}$, $\bar{X}$, $\bar{Z}$ and $\bar{U}$ in Fig. \ref{fig1}(b) possess the wavevector groups with non-trivial representations.


The uniqueness of a non-symmorphic group lies in the structure of its representations\cite{bir1974}. Although the factor group of the translation subgroup in a non-symmorphic group is isomorphic to a point group, its representation is not identical to the conventional representation of a point group.
For an element $\{S|\vec{R}\}$ in a space group, the corresponding representation matrix at a momentum $\vec{k}$ takes the form $D_{\vec{k}}(\{S|\vec{R}\})=e^{i\vec{k}\cdot\vec{R}}D(S)$.
The matrix $D$, that only depends on the point group operation, satisfies the multiplication rule $D(S_1)D(S_2)=\omega(S_1,S_2)D(S_1S_2)$ for two symmetry operators $\{S_1|\vec{R}_1\}$ and $\{S_2|\vec{R}_2\}$ in the wavevector group, where $\omega$ is a phase given by $\omega(S_1,S_2)=e^{i(\vec{k}-S_1^{-1}\vec{k})\cdot\vec{R}_2}$ and it defines a so-called factor system\cite{bir1974}. The additional phase coefficient appearing in the multiplication rules indicates that projective representations of a point group, instead of conventional representations, are required to describe a non-symmorphic symmetry group. The projective representations are usually classifed into different classes by their factor systems. To determine the class for a wavevector group, one can consider the parameter $\alpha=\omega(S_1,S_2)/\omega(S_2,S_1)$, where $S_1$ commutes with $S_2$. For a crystalline symmetry group, $\alpha$ can only take values of $\pm 1$. If $\alpha=1$, the projective representation belongs to a class identical to the conventional representation, denoted as $K_0$. If $\alpha=-1$, the projective representation belongs to a non-trivial class, usually denoted as $K_1$.

We may consider our example of pmg group. The corresponding factor group is isomorphic to the $D_2$ group. The $K_0$ class is the same as the conventional representation, which only contains 1D IRs, while the $K_1$ class of $D_2$ group has one 2D IRs, which indicates the double degeneracy at HSM. In the surface BZ (Fig. \ref{fig1}(b)), only four momenta $\vec{K}=\bar{\Gamma},\bar{X},\bar{Z},\bar{U}$ contain all the symmetry operations in pmg group. For the operators $\{\hat{m_z}|\vec{e}\}$ and $\hat{g}_x=\{\hat{m}_x|\vec{\tau}\}$, we have $\alpha=\omega(\hat{m}_z,\hat{m}_x)/\omega(\hat{m}_x,\hat{m}_z)=e^{i\vec{\tau}\cdot(\hat{m}_x^{-1}\vec{K}-\hat{m}_z^{-1}\vec{K})}$. Direct calculation shows $\alpha=1$ for $\vec{K}=\bar{\Gamma},\bar{X}$ and $\alpha=-1$ for $\vec{K}=\bar{Z},\bar{U}$. Therefore, all the states at $\bar{Z}$ and $\bar{U}$ must be doubly degenerate, consistent with the analysis of our concrete tight-binding model.

The analysis based on projective representations can be applied to surfaces with other 2D non-symmorphic groups, namely the pg, pgg and p4g groups. The typical surface lattice structures for pgg and p4g groups are shown in Fig. \ref{fig3}(c) and (d), respectively. Here we always consider the surface normal to y direction and assume the lattice along the y direction preserves the 2D symmetry group of the surface. Thus, the degeneracies found in the surface BZ are preserved along the whole $k_y$ line in the 3D bulk BZ, which allows us to define a $Z_2$ topological invariant in a way similar to the case of pmg group.
The classes of the projective representations for HSM $\bar{\Gamma},\bar{X},\bar{Z},\bar{U}$ are summarized in Table \ref{tab1} for different non-symmorphic groups. One finds no $Z_2$ topological phases in pg group since all HSM belong to the $K_0$ class. For both the pgg and p4g groups, $\bar{X}$ and $\bar{Z}$ belong to the $K_1$ class, so topological surface states can exist between $\bar{X}$ and $\bar{Z}$. For the p4g group, $\bar{\Gamma}$ and $\bar{U}$ bleong to the $K_0$ class of the $D_4$ group, which contains two 1D IRs and one 2D IR. Therefore, both doublets and singlets exist at these two momenta, similar to the case of the P4m group\cite{alexandradinata2013}. The complete study of topological phases in p4g group will be given elsewhere.

\begin{table}[htb]
  \centering
  \begin{minipage}[t]{1.\linewidth} 
      \caption{ The degeneracy of HSM in 2D non-symmorphic groups. Here ``FG'' is for the factor groups of the translation subgroup and ``Deg'' is for degeneracy. ``$Z_2$'' of the last column means whether $Z_2$ topological phases can exist in this non-symmorphic group.  }
\label{tab1}
\hspace{-1cm}
\begin{tabular}{|c|c|c|c|c|c|}
    \hline\hline
    Group & HSM & Class & FG & Deg & $Z_2$ \\
    \hline
    pg &  $\bar{\Gamma},\bar{X},\bar{Z},\bar{U}$ & $K_0$ & $D_1$ & 1 & No  \\
    \hline
    pmg & $ \begin{array}{c}
	    \bar{\Gamma},\bar{X}\\
            \bar{Z},\bar{U}
    \end{array}$  &  $ \begin{array}{c}
	    K_0\\
            K_1
    \end{array}$ & $ \begin{array}{c}
	    D_2\\
            D_2
    \end{array}$ & $ \begin{array}{c}
	    1\\
            2
    \end{array}$ & Yes  \\
    \hline
    pgg & $ \begin{array}{c}
	    \bar{\Gamma},\bar{U}\\
            \bar{X},\bar{Z}
    \end{array}$  &  $ \begin{array}{c}
	    K_0\\
            K_1
    \end{array}$ & $ \begin{array}{c}
	    D_2\\
            D_2
    \end{array}$ &  $ \begin{array}{c}
	    1\\
            2
    \end{array}$ & Yes  \\
    \hline
    p4g & $ \begin{array}{c}
	    \bar{\Gamma},\bar{U}\\
            \bar{X},\bar{Z}
    \end{array}$  &  $ \begin{array}{c}
	    K_0\\
            K_1
    \end{array}$ & $ \begin{array}{c}
	    D_4\\
            D_2
    \end{array}$ &  $ \begin{array}{c}
	    1 \mbox{ or } 2\\
            2
    \end{array}$ & Yes \\
    \hline
  \end{tabular}%
  \end{minipage}
\end{table}

{\it Discussion and conclusion -}
To realize TNCIs, one needs to look for semiconducting materials that possess surfaces with 2D symmetry groups pmg, pgg and p4g. It is known that 157 of the 230 space groups are non-symmorphic and the surfaces with the required symmetry should exist in many of them. For example, the surfaces with pmg group can exist in many compounds of iron pnictides and chalcognides ($P4/nmm$ group)\cite{kamihara2008,hu2013} and some II-V narrow gap semiconductors with $Cd_3As_2$ type of structures ($P4_2/nmc$ group)\cite{steigmann1968,wang2013}. A systematic search for TNCIs in these classes of materials is required.

We conclude our discussion with three comments. Firstly, in the TCI model proposed by Fu\cite{fu2011a}, the existence of singlets weakens the stability of surface states. For non-symmorphic groups, only doublets can exist at certain HSM and we expect topological surface states of TNCIs are more robust. Secondly, we focus on the single group cases, which can be directly applied to electron systems with no spin-orbit coupling, or to bosonic systems such as photonic crystals\cite{yannopapas2011,wang2009,rechtsman2013}. The generalization to double groups is straightforward and will be discussed elsewhere. Thirdly, the degeneracies due to non-symmorphic symmetry groups have been discussed in the context of interacting electrons recently\cite{roy2012,parameswaran2013}. Thus, it is interesting to find out if TNCIs can also exist for interacting electrons.


We would like to thank X. Dai, L. Fu, V. Gopalan, X.L. Qi, Y.H. Wu and B. VanLeeuwen for useful discussions. We especially acknowledge Y.H. Wu's careful reading of the manuscript. CXL was supported by startup funds from PSU. 

\appendix
\section{Tight-binding model of topological non-symmorphic crystalline insulators}

The tight-binding Hamiltonian of our model is given by
\begin{eqnarray}
	&&H=H_{A}+H_{B}+H_{AB},\label{eq:Ham1}\\
	&&H_{\eta}=\sum_{\langle \vec{n}\vec{m}\rangle_{in}, \alpha\beta}t_{\vec{n}\vec{m}}^{\alpha\beta}c^{\dag}_{\alpha \vec{n}\eta}c_{\beta \vec{m}\eta}+\sum_{\vec{n},\alpha}\epsilon_{\alpha}c^{\dag}_{\alpha \vec{n}\eta}c_{\alpha \vec{n}\eta}\nonumber\\
	&&+\sum_{\vec{n}}\delta_{\eta} M_1(-ic^{\dag}_{\vec{n}p_x\eta}c_{\vec{n}p_y\eta}+ic^{\dag}_{\vec{n}p_y\eta}c_{\vec{n}p_x\eta})\\
	&&H_{AB}=\sum_{\langle \vec{n}\vec{m}\rangle_{AB}, \alpha\beta}(r_{\vec{n}\vec{m}}^{\alpha\beta}c^{\dag}_{\alpha \vec{n}A}c_{\beta \vec{m}B}+h.c.)
\end{eqnarray}
where $\eta=A,B$ is for A,B layers, $\delta_{\eta=A(B)}=+(-)$, $\vec{n}=(n_x,n_y,n_z)$, $\vec{m}=(m_x,m_y,m_z)$ denote lattice sites and $\alpha,\beta=s,p_x,p_y$ denote orbitals. The term $H_A$ ($H_B$) comes from hopping terms within the layer consisting of only A (B) atoms, while $H_{AB}$ is due to hopping between A and B layers. $\langle nm\rangle_{in}$ represents nearest neighbors in the xy plane with hopping parameters $t^{\alpha\beta}_{\vec{n}\vec{m}}$ while $\langle \vec{n}\vec{m}\rangle_{AB}$ represents nearest neighbors that reside in two adjacent A and B layers with the parameters $r^{\alpha\beta}_{\vec{n}\vec{m}}$. We take into account the $\sigma$ bond for the s orbitals, the $\sigma$ and $\pi$ bonds for the p orbitals, and the $\sigma$ bonds between the s and p orbitals.
The intra-layer hopping parameters are given by the matrices
\begin{eqnarray}
	&&t_{\vec{n},\vec{n}+\hat{e}_x}=\left(
	\begin{array}{ccc}
		u_{s\sigma}&u_{sp\sigma}&0\\
		-u_{sp\sigma}&u_{p\sigma}&0\\
		0&0&u_{p\pi}
	\end{array}
	\right),\nonumber\\
        &&t_{\vec{n},\vec{n}+\hat{e}_y}=\left(
	\begin{array}{ccc}
		u_{s\sigma}&0&u_{sp\sigma}\\
		0&u_{p\pi}&0\\
		-u_{sp\sigma}&0&u_{p\sigma}
	\end{array}
	\right),
	\label{eq:parameter_inplane}
\end{eqnarray}
in the basis $|s\rangle$, $|p_x\rangle$ and $|p_y\rangle$, where $\hat{e}_x$ and $\hat{e}_y$ denote unit vectors to the nearest neighbor site along the x and y directions, respectively. For the hopping between two layers, since $\vec{r}_0=\vec{r}_B-\vec{r}_A$ is not along the z direction, we need to decompose the p orbitals into components along $\vec{r}_0$ axis and perpendicular to $\vec{r}_0$. Consequently, we obtain 
\begin{eqnarray}
	r_{\vec{n},\vec{n}+\hat{r}_0}=\left(
	\begin{array}{ccc}
		v_{s\sigma}&v_{sp\sigma}\lambda_1&0\\
		-v_{sp\sigma}\lambda_1&v_{p\sigma}\lambda_1^2+v_{p\sigma}(1-\lambda_1^2)&0\\
		0&0&v_{p\pi}
	\end{array}
	\right),
	\label{eq:parameter_outofplane}
\end{eqnarray}
where $\lambda_1=\frac{2a_1}{|\vec{r}_0|}$ is the angle between $\vec{r}_0$ and the x axis.
$M_1$ term is the Zeeman type of coupling between p orbitals and magnetic moments.
In momentum space, the Hamiltonian is given by
\begin{widetext}
\begin{eqnarray}
	H_{\eta}=\sum_k\Psi^{\dag}_{\eta}\left(
	\begin{array}{ccc}
		E_s(\vec{k})&-2iu_{sp\sigma}\sin(k_xa)&-2iu_{sp\sigma}\sin(k_ya)\\
		2iu_{sp\sigma}\sin(k_xa)&E_{px}(\vec{k})& -i\eta M_1\\
		2iu_{sp\sigma}\sin(k_ya)&i\eta M_1&E_{py}(\vec{k})
	\end{array}
	\right)\Psi_{\eta}
	\label{eq:HamA}
\end{eqnarray}
\begin{eqnarray}
	H_{AB}=\sum_k\Psi^{\dag}_{+}e^{i2k_x a_1}\left(
	\begin{array}{ccc}
		2v_{s\sigma}\cos(k_zc/2)&-2v_{sp\sigma}\lambda_1\cos(k_zc/2)&0\\
		2v_{sp\sigma}\lambda_1\cos(k_z/2)&2\cos(k_zc/2)(v_{p\pi}(1-\lambda_1^2)+v_{p\sigma}\lambda_1^2)&0\\
		0&0&2v_{p\pi}\cos(k_zc/2)
	\end{array}
	\right)\Psi_{-}
	\label{eq:HamAB}
\end{eqnarray}
\end{widetext}
where
\begin{eqnarray}
	&&E_s(\vec{k})=2u_{s\sigma}(\cos(k_xa)+\cos(k_ya))+\epsilon_s\nonumber\\
	&&E_{px}(\vec{k})=2(u_{p\sigma}\cos(k_xa)+u_{p\pi}\cos(k_ya))+\epsilon_p\nonumber\\
	&&E_{py}(\vec{k})=2(u_{p\pi}\cos(k_xa)+u_{p\sigma}\cos(k_ya))+\epsilon_p.
	\label{eq:HamEsEp}
\end{eqnarray}
Here $\Psi^\dag_{\eta}=(c^\dag_{s\eta}(\vec{k}),c^\dag_{p_x\eta}(\vec{k}),c^\dag_{p_y\eta}(\vec{k}))$, and $\vec{k}=\sum_{i=x,y,z}k_i\vec{e}_i$.

Next we would like to analyze the constraint on the form of Hamiltonian due to symmetry. In the main text, we have shown how the operations $\hat{m}_z$ and $\hat{g}_x$ act on the basis wavefunctions. The corresponding transformation matrices are given by
\begin{eqnarray}
	&&U(\hat{m}_z)=\left(
	\begin{array}{cc}
		1&0\\
		0&1
	\end{array}
	\right)\label{eq:Umz}\\
	&&U(\hat{g}_x)=\left(
	\begin{array}{cc}
		0&e^{-i(\hat{m}_x\vec{k})\cdot\vec{\tau}}m_x\\
		e^{-i(\hat{m}_x\vec{k})\cdot\vec{\tau}}m_x&0
	\end{array}
	\right)
	\label{eq:Ugx}
\end{eqnarray}
where
\begin{eqnarray}
	m_x=\left(
	\begin{array}{ccc}
		1&0&0\\
		0&-1&0\\
		0&0&1
	\end{array}
	\right).
\end{eqnarray}

A symmetry of a Hamiltonian requires 
\begin{eqnarray}
	H(\vec{k})=U^*(\vec{k})H(\hat{U}\vec{k})U^T(\vec{k}). 
	\label{eq:symmetry}
\end{eqnarray}
In particular, the mirror symmetry $\hat{m}_z$ yields
\begin{eqnarray}
	&&H_{A(B)}(k_x,k_y,k_z)=H_{A(B)}(k_x,k_y,-k_z),\label{eq:Hk_mz1}\\
	&&H_{AB}(k_x,k_y,k_z)=H_{AB}(k_x,k_y,-k_z). 
	\label{eq:Hk_mz2}
\end{eqnarray}
The glide symmetry results in
\begin{eqnarray}
	&&H_{A}(k_x,k_y,k_z)=m_xH_B(-k_x,k_y,k_z)m_x\label{eq:Hkg1}\\
	&&H_{AB}(k_x,k_y,k_z)=m_xH_{BA}(-k_x,k_y,k_z)m_x.
	\label{eq:Hkg2}
\end{eqnarray}
Moreover, since $|\alpha\eta,\vec{k}+\vec{G}\rangle=e^{i\vec{G}\cdot\vec{r}_{\eta}}|\alpha\eta,\vec{k}\rangle$, one find
\begin{eqnarray}
	&&H_{A(B)}(\vec{k}+\vec{G})=H_{A(B)}(\vec{k})\label{eq:HG1}\\
	&&H_{AB}(\vec{k}+\vec{G})
	=e^{i\vec{G}\cdot\vec{r}_{0}}H_{AB}(\vec{k})
	\label{eq:HG2}
\end{eqnarray}
with $\vec{r}_{0}=(2a_1,0,\frac{c}{2})$.
Equations (\ref{eq:Hk_mz1})-(\ref{eq:HG2}) determine the form of our tight-binding Hamiltonian that respects the pmg symmetry group.

For each individual layer, the Hamiltonian $H_{\eta}$ is the same as that in Ref. \cite{liu2013c}, where it was shown that, with proper choices of parameters, the low energy physics of each layer can be described by a quantum anomalous Hall model. The band gap of the quantum anomalous Hall model can be tuned by Zeeman type of couling $M_1$. When $M_1$ exceeds a critical value, the transition between trivial and non-trivial states can happen. This transition just corresponds to the transition between $Z_2$ trivial and non-trivial phases in our model. In the following, we will numerically calculate the energy dispersion in a slab geometry with the open boundary condition along the y direction to test the existence of gapless surface states, We choose a set of parameters as:
$u_{s\sigma}=-0.2$, $u_{sp\sigma}=0.2$, $u_{p\sigma}=0.2$, $u_{p\pi}=0.2$, $v_{s\sigma}=0.05$, $v_{sp\sigma}=0.3$, $v_{p\sigma}=-0.1$, $v_{p\pi}=0.1$, $\epsilon_s=0$, $\epsilon=-5$, $a=1$, $a_1=0.1$ and $c=2$. Combining both slab calculations and bulk dispersions, we find band gap closes at three different $M_1$ values. These band gap closings correspond to three (in fact two) topological phase transitions: (I) $M_1=3.4$, the system changes from being topological trivial (no gapless surface state) to topological non-trivial (single gapless surface state at $\bar{Z}$). (II) $M_1=5$, the system remains topological non-trivial since band gap closes at both $T$ and $U$ at the same time, similar to the case discussed in Ref. \cite{slager2012}. So the gapless surface states that appear at $\bar{Z}$ previously moves to $\bar{U}$. (III) $M_1=6.6$, the system changes from being topological non-trivial to topological trivial.
\begin{figure}
   \begin{center}
      \includegraphics[width=3.5in,angle=0]{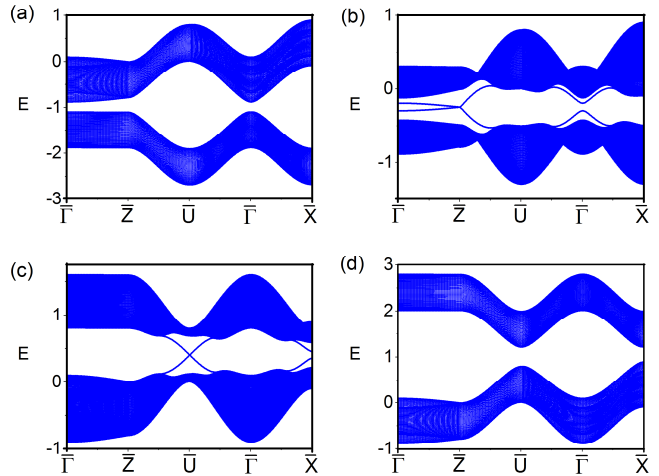}
    \end{center}
    \caption{ Energy dispersion with a slab configuration at different $M_1$ values: (a)$M_1=3.1<3.4$, system is topological trivial. (b) $3.4<M_1=4.5<5$, system is topological non-trivial with single gapless surface state at $Z$. (c) $5<M_1=5.8<6.6$, system is topological non-trivial with single gapless surface state at $U$ instead of $Z$. (d) $M_1=7>6$, system is topological trivial. }
    \label{fig4}
\end{figure}

\begin{figure}
   \begin{center}
      \includegraphics[width=3.5in,angle=0]{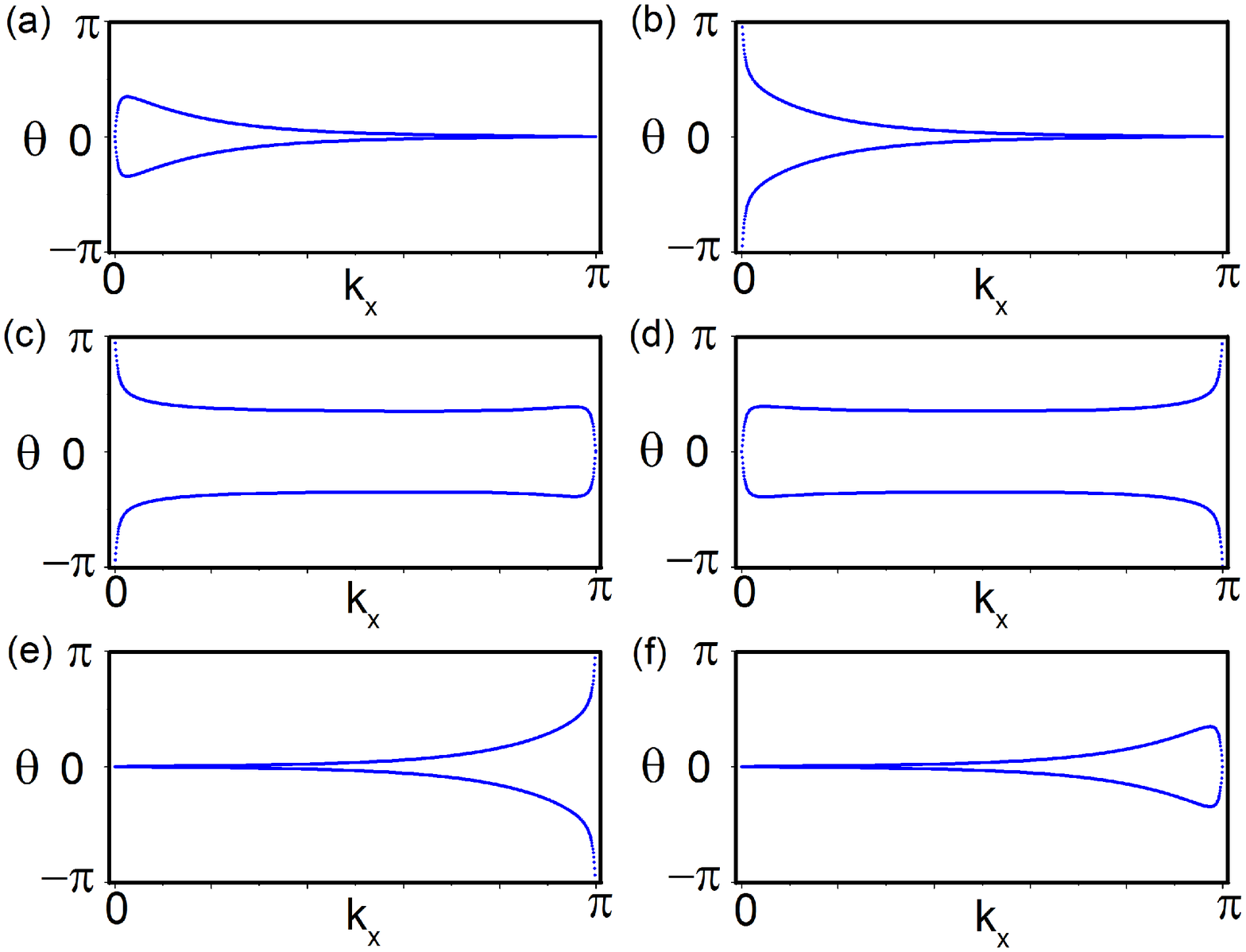}
    \end{center}
    \caption{ Wannier center flows for our tight-binding model characterizing three topological phase transitions (TPT): First TPT at $M_1=3.4$ : (a) $M_1=3.39$ (winding number=0); (b) $M_1=3.41$ (winding number=1). Second TPT at $M_1=5$ (this is a fake TPT since band gap closes twice at the same time): (c) $M_1=4.99$ (winding number=1); (d) $M_1=5.01$ (winding number=1). Third TPT at $M_1=6.6$ : (e) $M_1=6.59$ (winding number=1); (f) $M_1=6.61$ (winding number=0). }
    \label{fig5}
\end{figure}

Besides direct calculations of surface states, we can also extract the bulk topological invariant by tracking the evolution of Wannier function centers in the $k_z=\pi/c$ plane for our tight-binding model. For each fixed $k_x$, Wannier function centers can be obtained using a gauge-independent method introduced by Yu {\it et. al.}\cite{yu2011}. For a one-dimensional system with periodic boundary conditions, the position operator is defined as,
\begin{equation}
\hat{X}=\sum_{i,\alpha}e^{-i\frac{2\pi}{L}\cdot {\bf R}_{i} }|\alpha,i\rangle \langle\alpha,i|
\end{equation}
where $L=N_{y} a$ is the length of the system, $\alpha$ is the orbital index and $i$ labels the lattice site. This position operator is defined using local basis $|\alpha, i\rangle$, so its eigenvalues represent Wannier function centers of this system. By projecting this position operator into the occupied bands, it is easy to check that the projected position operator is equivalent to a $U(2N)$ Wilson loop for fixed $k_x$,
 \begin{equation}
 D(k_x)=S_{0,1}S_{1,2}S_{2,3}...S_{N_y-2,N_y-1}S_{N_y-1,0}
 \end{equation}
 where a series of overlap matrices $S$ are defined using the periodic parts of Bloch wave-functions
 \begin{eqnarray}
 S_{i,i+1}^{m,n}(k_x)& = & \langle m, k_{y,i},k_x|n, k_{y,i+1}, k_x\rangle \nonumber\\
 k_{y,i} & = & \frac{2\pi i}{N_{y} a}
 \end{eqnarray}
 Then the phase of the eigenvalues of this $U(2N)$ Wilson loop $D(k_x)$ just give us Wannier centers of the occupied bands. Let $k_x$ evolves from $0$ to $\pi$, we could clearly see whether the Wannier centers switch partners (when the winding number is odd and the system is topologically trivial) or not (when the winding number is even and the system is topologically trivial). As is shown in both Fig. \ref{fig4} and Fig. \ref{fig5}, the winding numbers of Wannier function centers precisely characterize topological phase transitions and the appearance of surface states.


\end{document}